# Arbitrage opportunities in publication and ghost authors

Lawrence Smolinsky

Department of Mathematics, Louisiana State University, Baton Rouge, LA USA
(smolinsk@math.lsu.edu)

**Abstract.**

In some research evaluation systems, credit awarded to an article depends on the number of co-authors on the article, with total credit to the article increasing with the number of co-authors. There are many examples of such evaluation systems (e.g., the United States National Research Council evaluation of graduate programs gave full credit to each co-author). Such credit systems run the risk of encouraging ghost or honorary authorships. In a recent article, António Osório and Lutz Bornmann (2019) propose a scheme to discourage ghost authorships but increase the total credit to a paper when co-authorships increase. This article shows that if articles are valued more highly as the number of co-authorships increases, then there are opportunities to increase credit by mutually agreeing to add each other as authors. Unrelated authors of unrelated papers may all benefit by expanding their co-author list. I call this phenomena arbitrage—a term borrowed from economics and finance—since the content of the articles do not change, but the value increases by moving to a "market" of more co-authors where articles are valued differently.

## 1. Introduction

The issue of how to count or credit publications with several authors is an important question that is addressed in the bibliometric and research evaluation literature. In a recent Journal of Informetrics article, António Osório and Lutz Bornmann (2019), point out that assigning the same credit to an article regardless of the number of co-authors "ignores the potential synergies that result from collaborations" (p. 541). They recognize the possibility that papers with more co-authors have more value and include a theoretical study.

Osório and Bornmann are aware of the potential for encouraging ghost or honorary authorships. They addressed the issue of creating "distortions in comparing individuals with different co-authorship patterns" and "incentives to the addition of 'ghost' co-authors, which is not desirable." They stated that "this issue is the objective of this paper" (p. 541). Their credit system does give a disincentive to adding ghost authors to a single publication. However, it is shown in this article, there are always incentives in any credit system for thoughtful authors to act in concert when the total credit or value of a publication increases with the number of authors. This correspondence gives a theoretical property of credit systems but is not a practical guide to increase credit through ghost authorships.

## 2. Background and notation

The notions in Osório and Bornmann (2019) are individual effort, credit, value of a collaboration, and the value when successful (published or accepted). Egghe, Rousseau, and Van Hooydonk (2000) and more recently Osório (2018) also address various credit systems. I use the notation from Osório (2018), since Osório and Bornmann (2019) assume equal credit to all authors on an article.

A credit and value system may be a counting system such as those discussed by Egghe, Rousseau, and Van Hooydonk (2000) or may instead be a predetermined function of the citation count accumulating after publication. In the second case, credit is a random variable. Credit to the paper as a whole will be examined as dependent on $n$, the number of co-authors, to give credit $c^n$, but that is not to say that other factors may not affect the value. A particular co-author's position $i$ is assumed to determine his or her share of the credit, $c_i^n$, of the total credit. The individual credit $c_i^n$ may be independent of $i$ as Osório and Bornmann assume, but I do not make that assumption.

Assume the principle that the credit awarded to a paper is the sum of individual credits to each author on the paper, or

A.1  $c^n = c_1^n + c_2^n + \cdots + c_n^n.$

Also assume that credit is non-negative, $c_i^n \geq 0$.

A ghost author is a co-author, whose contribution does not warrant an authorship recognition. If a ghost author is added to an article, then the article's total value goes from $c^n$ to $c^{n+1}$. The decision to add ghost authors must be made prior to publication and the accumulation of citations. This means the decision to be measured in the market of $n$ authors $c^n$ or $n+1$ authors $c^{n+1}$ is made prior to publication. In the scenario that credit is awarded after publication as a random variable by citation counts, I assume the credit or value $c^n$ is taken to be the expected value of the random variable based on the citation count. In the Osório and Bornmann notation, $c^n = \bar{v}_n$. Hence, $c^n$ and $c_i^n$ are numerical for the remainder of this article.

There is evidence that citation counts of publications increase with the number of co-authors or

A.2  $c^n < c^{n+1}.$

Onodera and Yoshikane (2015) give a summary of studies considering various citation-influencing factors (Table 1, p. 743). They concluded that five studies found the number of coauthors a "strong or definite predictor"; four studies found the number of coauthors a "weak predictor or predictive power dependent on the model"; and four studies found the number of coauthors "not significant or negative predictor." Citing Onodera and Yoshikane and others, Osório and Bornmann accept that adding more authors should increase the expected value of a publication and so A.2 holds. It may seem reasonable that visibility and the citation count improve if there are more co-authors (ghost or not) as there may be more close connections to the article's authors. They may have students or former students, past or present collaborators, colleagues, and more "team self-citations" (Garfield 1979:245).

Note that A.2 may apply to research evaluation even if one does not wish to consider the citation evidence. An important example comes from the United States National Academies. Their operating arm, the National Research Council, periodically evaluates graduate programs in the United States. When counting publications per faculty, they counted $c^n = n$ so that each co-author receives credit for a full article (National Research Council. 2011). This is called "total author counting" (Egghe et al. 2000, p. 146).

## 3. Arbitrage opportunities

It is true that an evaluation system will have a disincentive to adding ghost authors to a single publication, if $c_{i+j}^{n+1} < c_i^n$ holds for $j \geq 0$. The *i*-th author will lose credit if an additional author is added to the list. However, this conclusion does not extend to more complicated arrangements where multiple authors are willing to add ghost authors to their publications in exchange for becoming ghost authors on other publications.

Imagine the simplest situation. Authors A and B are preparing article I and article II respectively. If the authors separately publish the two articles as single authors, then each author will receive $c^1$ credit. If they combine, then they may add the other author as a second author. Article I will have authors A & B and article II will have authors B & A. Yes, author A may get less credit from article I, only $c_1^2$. However, author A will get more total credit from the two articles, $c_1^2 + c_2^2 = c^2$ credit, since $c^2 > c^1$. There may be a no-arbitrage principle in financial markets, but there are arbitrage opportunities in academic markets under A.2. If a publication has greater value in the market with more coauthors, then move to that market.

There are always academic arbitrage opportunities for authors or research groups to profit without additional effort by ghost author exchanges. The example of two single authors is an example of two research groups each consisting of a single person. It may be generalized to an arbitrary number and sizes of groups by a theoretical argument. I do not try to control the credit within each group, as the group may wish to control that assignment.

**Proposition**: Consider a non-negative credit system where A.1 holds. If papers with more co-authors are worth more total credit than papers with fewer co-authors (i.e., A.2 holds), then for any collection of research groups of any size whose membership is distinct, there are opportunities to increase the publication credit for all research groups by an exchange of ghost authors.

The proof is given in the appendix.

Some have considered ghost authorships outright fraud as discussed by Blaise Cronin (2005). Nevertheless, nearly any contribution to an article may constitute justification as a co-author and generosity in authorship recognition has not been considered immoral. It may be hard to distinguish ghost authorship from minimal contributions. Furthermore, if there is a desire among research evaluators to discourage authorship exchanges, then it is not clear who the gatekeeper of co-authorship restraint is to be:
- Ghost authorship exchanges benefit researchers by increasing their individual credit.

- Ghost authorship exchanges benefit universities and research institutes by increasing the aggregate credit awarded to the institution. The institution's bibliometric evaluation improves.
- If expected values of citation counts do increase with the number of co-authors, then journal editors also benefit from more co-authors per paper, e.g., to increase journal impact factors.

If ghost authorship exchanges are to the benefit of researchers, administrators, and editors, then it may be difficult to find an effective gatekeeper. It is a conundrum that seems to be fundamental to bibliometric counting for the purpose of convenient research evaluation.

## Acknowledgments

I wish to thank Aaron Lercher and Andrew McDaniel who read and commented on this article and particularly to Andrew McDaniel who reviewed the mathematics in the appendix.

## Appendix

To prove the Proposition, a ghost exchange strategy is introduced that increases the credit for each research group involved. It is theoretical in that the strategy involves a large number of papers and generally not practical. In many situations, there are easier ways for arbitrage gains that depend on the particular system of research evaluation.

For an example consider the total author counting system. Suppose there are $m$ research groups $\mathcal{A}_i$ for $i=1, \cdots, m$ and each can produce one paper for a total of $m$ papers. All the researchers and groups will gain if all the authors become authors on all papers rather than publishing the $m$ papers each with only the correct group of legitimate authors. The credit with legitimate authors is 1 for each author, but with the exchange of ghost authors, each author gets credit for $m$.

For the general case, the proposition is proved in a technical lemma below. The lemma gives the general method of construction that works for a credit scheme.

Lemma: Suppose there are $m$ research groups $\mathcal{A}_i$ for $i=1, \cdots, m$; $\mathcal{A}_i$ consists of $n_i$ researchers; and $\mathcal{A}_i \cap \mathcal{A}_j = \emptyset$ for $i \neq j$. Suppose the authors in each group receive credit by a non-negative credit system that satisfies A.1 and A.2. Then there are counting numbers $k_i$ for $i=1, \cdots, m$ with $k = k_1 + \cdots + k_m$, and
(1) If group $\mathcal{A}_i$ publishes $k_i$ papers with the (legitimate) authors $\mathcal{A}_i$, then the group $\mathcal{A}_i$ receives a total of $k_i c^{n_i}$ credit.
(2) If the groups publish all $k$ articles together in a ghost authorship exchange, then there are author orderings so each group $\mathcal{A}_i$ receives a total of $k_i c^n$ credit.

Since $k_i c^n > k_i c^{n_i}$, the credit is greater for each group by engaging the ghost authorship exchange.

Proof: Write $n = n_1 + n_2 + \cdots + n_m$, so that $n$ is the total number of researchers as the groups have distinct researchers. Pick $k_i$ and $k$ to be the multinomial coefficients):

$$k_i = \binom{n-1}{n_1,\ n_2,\ \cdots, n_i-1, \cdots,\ n_m}$$

and

$$k = \binom{n}{n_1,\ n_2,\ \cdots,\ n_m}$$

(Ghahramani, 2005). Note that $k = k_1 + \cdots + k_m$ by the multinomial version of Pascal's recursion formula (Abramowitz & Stegun, 1972, p. 823).

Each group $\mathcal{A}_i$ has $k_i$ papers for $i=1, \cdots, m$. The credit to a single paper published by the group of $n_i$ authors is $c^{n_i}$. Since the group $\mathcal{A}_i$ has $k_i$ papers, it will receive $k_i c^{n_i}$ total credit. Hence part (1) follows.

To address part (2), the notion of a permutation is required. A permutation is an ordered arrangement of objects (Ghahramani 2005). For example, the permutations of A, B, and C are six: ABC, ACB, BAC, BCA, CAB, and CBA. If two objects of the three are indistinguishable (or considered alike and for some purpose indistinguishable), then the distinguishable permutations are smaller in number. For example, the distinguishable permutations of A, A, and C are three: AAC, ACA, and CAA. Here the B is considered indistinguishable or like A and is denoted as a second A. It is a general counting principle that the number of distinguishable permutations of $n$ objects of $m$ different types, where $n_1$ are alike, $n_2$ are alike, . . . , $n_m$ are alike and $n = n_1 + n_2 + \cdots + n_m$, is $k = \binom{n}{n_1, n_2, \cdots, n_m}$, the multinomial coefficient (Ghahramani 2005).

Treat the $n_i$ members of the research group $\mathcal{A}_i$ as indistinguishable or alike. There are a total of $n$ objects (authors) separated into $m$ different types (research groups). A permutation of the $n$ objects is an ordering (like an author list on an article). Each of the $k$ articles will include all $n$ authors following the various ordering specified by each of the $k$ distinguishable permutations, which give the author orderings required in part (2). Note that if a particular place on one of the articles is supposed to go to a member of $\mathcal{A}_i$, then the construction does not specify which particular member of $\mathcal{A}_i$ it should be. The decision is left to the group $\mathcal{A}_i$ to decide.

The next issue is to show the total credit awarded to the group $\mathcal{A}_i$ from the collection of $n$ articles is $k_i c^n$. In how many permutations does a researcher from $\mathcal{A}_i$ appear in the first position? If the first position is filled by a member of $\mathcal{A}_i$, then other $n$-1 positions (that is positions 2 through $n$) may be filled by remaining researchers: $n_1$ from $\mathcal{A}_1$, $n_2$ from $\mathcal{A}_2$, etc. but only $n_i$-1 from $\mathcal{A}_i$ because one member of $\mathcal{A}_i$ is in the first position. By the counting principle, a member of $\mathcal{A}_i$ will appear in the first position $k_i$ times. Looking only at the first position, the group $\mathcal{A}_i$ receives $k_i c_1^n$ credit. There is nothing special about the first position and the argument applies to each of the $n$ positions. Hence the total credit awarded to the group $\mathcal{A}_i$ is

$$k_i(c_1^n + \cdots + c_n^n) = k_i c^n.$$

This completes the proof. QED

The argument may be changed to run through all the permutations of authors rather than treating the members of a group as indistinguishable. It would change $k$ and $k_i$ to $k = n!$ and $k_i = (n-1)!\, n_i$. This variation of argument would still improve the credit awarded to group $\mathcal{A}_i$ from $k_i c^{n_i}$ to $k_i c^n$. However, each author in group $\mathcal{A}_i$ would then get equal credit.